## Astronomy

# The Phoenix stellar stream rose from the ashes of an ancient star cluster

**J. M. Diederik Kruijssen**

Observations of an ancient stellar stream provide the first evidence of a vanished population of extremely metal-poor stellar clusters. Their remnants might reveal how the early assembly of the Milky Way proceeded.

Globular clusters are among the most enigmatic objects in the Universe. They are stellar systems of about 100,000 stars packed within just a few tens of light years. Most of them are almost as old as the Universe itself. Globular clusters are found orbiting nearly all known galaxies with more than a billion stars[1]. Their spheroidal distribution around these galaxies suggests that many globular clusters formed in other, low-mass galaxies that were since accreted by the central galaxy and were shredded by its tidal forces. In recent years, astrophysicists have learned to use globular clusters as fossils to reconstruct this galaxy assembly process[2]. However, globular clusters can be destroyed by the host galaxy's tidal forces too, in the process erasing the secrets they hold. On page 768, Wan *et al.*[3] report spectroscopic observations of a stellar stream that is the remnant of the most metal-poor globular cluster discovered to date, providing a unique perspective on the earliest epochs of galaxy assembly.

The chemical composition of the stars in a globular cluster is a key observable linking them to their natal galaxies. The stars in a cluster were born together, within the same parent molecular gas cloud, and have highly similar chemical compositions[4]. Specifically, all stars in a cluster share the same iron content, which in turn reflects the iron content of the galaxy in which they formed. Throughout cosmic history, iron is produced in supernova explosions, which chemically enrich the gas consumed in future generations of star formation. The enrichment cycle proceeds more rapidly in galaxies with higher masses and star formation rates, so that the iron content (or "metallicity") increases not only with time, but also with galaxy mass[5]. This galaxy mass-metallicity relation is predicted to evolve quite slowly in the early history of the Universe[6], when globular clusters formed. As a result, the metallicity of a globular cluster indicates the mass of the galaxy in which it formed.

Due to the galaxy mass-metallicity relation, there must be a metallicity below which early-Universe galaxies contain fewer stars than typical globular clusters do, suggesting that they might not have been able to form such massive clusters. Theoretical models predict that this transition happens at metallicities of about 0.3% of the metallicity of the Sun[7]. At this metallicity, early-Universe galaxies should have contained just a few 100,000 stars, very similar to the number of stars in typical, present-day globular clusters. This need not be a sharp transition, because the galaxy mass-metallicity relation exhibits considerable scatter, but observations do indeed reveal a dearth of globular clusters with metallicities less than 0.3% solar[8]. Lower-metallicity globular clusters are predicted to have existed, but they were necessarily less massive and less bound by gravity. As a result, these extremely metal-poor stellar clusters would have been destroyed by tidal forces from their host galaxy over cosmic time[7]. If this hypothesis is correct, the remnants of extremely metal-poor globular clusters might still be orbiting the Milky Way.

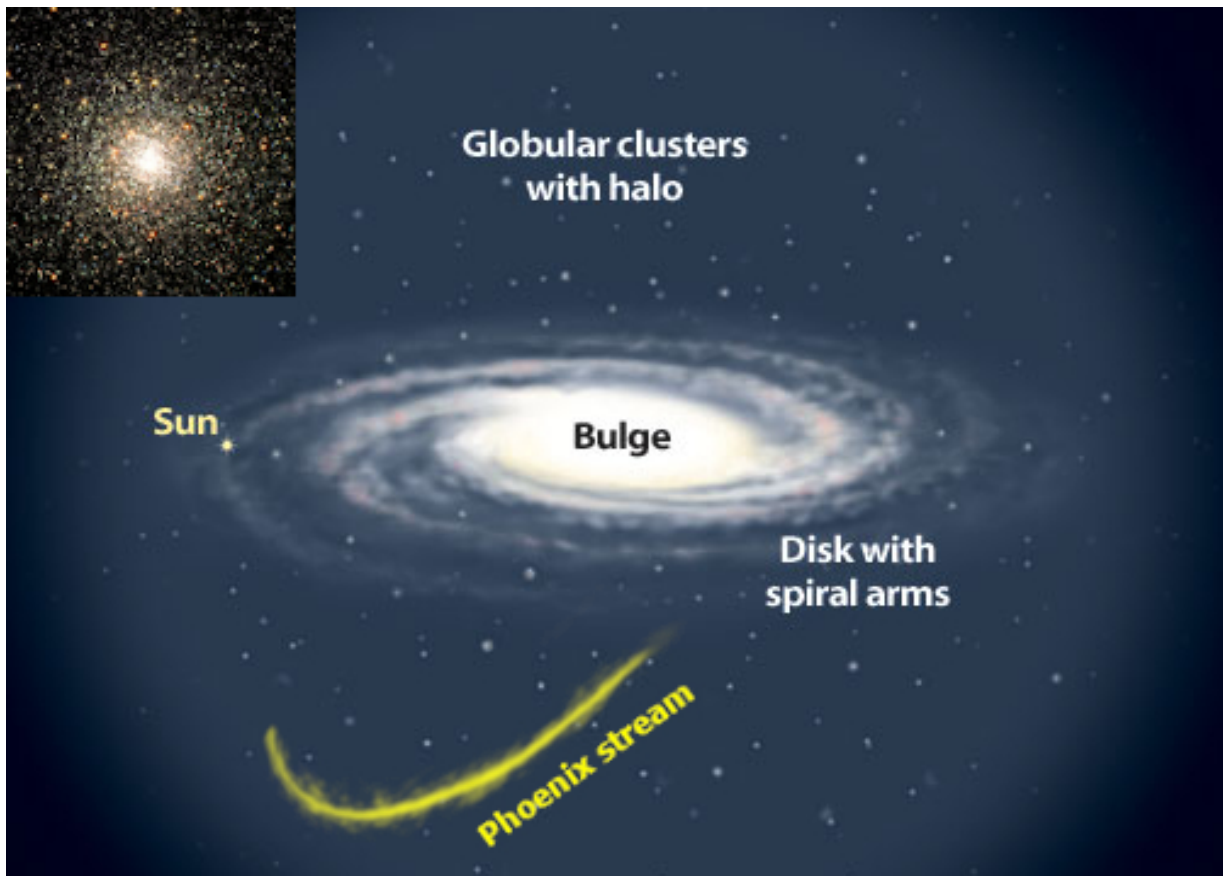

**Figure 1 | The extremely metal-poor Phoenix stream orbiting the Milky Way.** Galaxies like the Milky Way are surrounded by more than a hundred globular clusters (visible as dots; also see inset). Wan *et al.*[3] report that the Phoenix stellar stream (visible in yellow) is a tidally disrupted globular cluster that is significantly more metal-poor than any surviving globular cluster in the Milky Way. This confirms theoretical predictions that the Milky Way should host the debris of a vanished population of such extremely metal-poor clusters, which formed in the early history of the Universe. (Background image credit: Sky & Telescope, 2003)

Wan *et al.* used the observations of the Southern Stellar Streams Spectroscopic Survey ($S^5$) to measure the metallicities of 11 stars in the Phoenix stellar stream, a group of stars orbiting the Galactic Centre at a distance of about 60,000 light years[9]. Surprisingly, the researchers measured extremely weak absorption lines in the stellar spectra, requiring a metallicity of just 0.20±0.03% of that of the Sun. Not only are the metallicities that low, but they are uniformly so, with a star-to-star spread similar to the measurement uncertainty (and much lower than metallicity spreads observed in dwarf galaxies). This means that the stars in the Phoenix stream were born in the same stellar cluster. Their unusually low metallicity has the exciting implication that the cluster must have formed at a time that its natal galaxy was one of the very lowest-mass galaxies.

Like all important discoveries, the measured metallicity of the Phoenix stream generates more questions than it provides answers. Although it is only a single object, it represents the first direct evidence that the Milky Way once hosted a vanished population of extremely metal-poor globular clusters. How numerous were these clusters? The discovery of more such remnants would herald a new and exciting way of reconstructing the demographics of the lowest-mass galaxy population from which the Milky Way assembled. By estimating the initial masses of such extremely metal-poor globular clusters, future studies could potentially determine what fraction of their natal galaxy's mass these clusters constituted, revealing how the lowest-mass galaxies formed and evolved in the early Universe. Direct observations of star-forming proto-galaxies in the early Universe with the upcoming

*James Webb Space Telescope* will be able to independently test the results of such studies[10]. Finally, by comparing the orbital kinematics of fossil stellar streams to those of groups of globular clusters believed to have arrived in the Milky Way during the same accretion event[11], it might be possible to assign them to specific progenitor galaxies in the Milky Way's ancestry.

Thanks to all-sky surveys reaching very low surface brightness and obtaining exquisite stellar kinematics, there has been a surge in the discovery of fossil stellar streams[12], many of which likely represent the remnants of tidally disrupted globular clusters. The discovery by Wan and colleagues makes it a priority to obtain accurate metallicities for all of these streams. Who knows how many relics like the Phoenix stream might be hiding in the Milky Way's halo? Now that the first has been found, the hunt is on.

**J. M. Diederik Kruijssen** is at the Astronomisches Rechen-Institut, Zentrum für Astronomie der Universität Heidelberg, Mönchhofstraße 12-14, 69120 Heidelberg, Germany.
e-mail: kruijssen@uni-heidelberg.de